# Order in disorder: increased carrier mobility of downscaled amorphous semiconductors for high-speed thin film transistors in flexible electronics

Yuezhou Luo, and Andrew John Flewitt

Electrical Engineering Division, Department of Engineering, University of Cambridge, Cambridge CB3 0FA, United Kingdom

Amorphous semiconductors are important channel semiconductors in thin film transistors (TFTs) which serve not only active-matrix displays, but also flexible electronics for Internet of Things (IoT) applications. Nevertheless, a great limitation of amorphous semiconductors is their low carrier mobilities relative to their monocrystalline counterparts. Based on a recently established band fluctuation framework [Y. Luo and A. Flewitt, *Phys. Rev. B* **109**, 104203 (2024)], this paper shows that the intrinsic carrier mobility of amorphous semiconductors can significantly increase simply through device downscaling, without any material-level optimizations. Specifically, it is revealed that the intrinsic electron mobility of hydrogenated amorphous silicon in a 10-nm long gap can increase by ∼ 12 times, and does not compromise device-to-device uniformity. This mobility improvement is a result of reduced localized band tail states due to the ultra-short gap length relative to the band fluctuation length scale before downscaling; the latter is determined by the short- and medium-range structural order of the amorphous semiconductor.

## I. INTRODUCTION

Thin film transistors (TFTs), traditionally known as the building blocks of active-matrix displays, now play an increasingly important role in Internet of Things (IoT) hardware that is seamlessly integrating into the objects in our daily life [1]. In contrast to silicon-based metal oxide semiconductor field effect transistors (MOSFETs) devices used in mature integrated circuits (ICs), many of the fascinating merits that TFTs have offered to IoT, such as large-area low-cost fabrication and low thermal budget, result from the property and fabrication process of their channel semiconductors which are typically non-crystalline materials [2,3], with amorphous semiconductors being the most commercially influential representatives. Well-known examples include hydrogenated amorphous silicon (*a*-Si:H) which demonstrated its success in early liquid crystal displays (LCDs) in the last century [4], as well as amorphous oxides (such as amorphous indium gallium zinc oxide, *a*-IGZO) which have been increasingly dominant since 2004 [5]. Speed, not a top priority for TFTs in the past, is now of growing significance in light of the soaring development of IoT technologies exemplified by flexible electronics [6,7]; this is especially so when it comes to emerging fields such as flexible microprocessor [8], flexible machine learning engine [9], and novel memory architecture for edge computing [10]. Amorphous semiconductors, however, are intrinsically disadvantaged in speed due to their low carrier mobility. For instance, the electron mobility of *a*-Si:H is typically limited to ∼ 1 cm²/(V s)

[11]. Even for the amorphous oxides which exhibit higher mobilities [such as ∼ 10 cm²/(V s) mobility of *a*-IGZO [5]], their mobility values are still very far from the expectations in standard IC electronics. Other material options available for TFTs, such as high-mobility polycrystalline and crystalline thin films, are at the cost of poor device uniformity, higher expenses, higher thermal budget and process complexity.

Great efforts have been made to improve the mobility of amorphous semiconductors. Some endeavors focused on the material level in order to improve the intrinsic mobility. These, for *a*-Si:H, include the use of disilane with a controlled flow rate [12], the technique of expanding thermal plasma [13], and the process of catalytic chemical vapor deposition [14]. For amorphous oxides such as *a*-IGZO, increase of carrier concentration through elemental controls [15,16], as well as decrease of defect density through metal capping [17] were attempted. Other endeavors focused on the device level, and aimed at improving the apparent field effect mobility through, for instance, optimizing semiconductor-insulator interface conditions [18-23] and decreasing the parasitic resistance [24]. Nevertheless, none of the reported methods so far presents an effective and reproducible way to fundamentally overcome the mobility issue; also, many of them often compromise other material merits such as stability, thermal budget and process simplicity. A new strategy is thus needed in light of the technological importance of amorphous semiconductors in the era of IoT.

One well-known factor causing the low mobilities of amorphous semiconductors is the localized states in the density of states (DOS) distribution band tails; this has been elaborated in a recent work of us [25], henceforth named "Paper 1". In Paper 1, we used $a$-Si:H as an example, and reinvestigated the essence of localized band tail states from the perspective of excess valence band (VB) and conduction band (CB) charges (termed "excess delocalized charges"), which induce extra potentials that force the crystalline silicon ($c$-Si) band edge to fluctuate. Fig. 1(a) shows the band edge fluctuation modeled in Paper 1, from which the fluctuation along a representative direction is extracted and shown in Fig. 1(b). As revealed in Paper 1, because of the existence of short- and medium-range structural order in amorphous semiconductors, the band fluctuation is slow. Here, Fourier decomposition in Fig. 1(c) reveals that the predominant (> 95%) Fourier components are below a cutoff spatial frequency of $0.012a_0^{-1}$, where $a_0$ is the short-range order length (3 Å) defined in Paper 1. This corresponds to a length scale of 25 nm. We therefore envisage that, similar to a high pass filter, downscaling below such a length scale would selectively retain only the high-frequency fluctuations with low amplitudes, therefore mitigating the overall band fluctuation. $a$-Si:H would then contain fewer localized band tail states according to the analysis in Paper 1, and thus exhibit a higher mobility.

The above hypothesis of band tail engineering by means of downscaling meanwhile also meets the general trend in the semiconductor industry; beside ICs, downscaling has also received much attention in TFTs. Nevertheless, existing studies did not straightforwardly corroborate our hypothesis above (e.g., Refs. [26-28]). This is due to two facts. First, it was the apparent field effect mobility (rather than the intrinsic mobility) that was investigated; the former is affected by non-ideality factors such as contact resistance. Second, the staggered TFT structures were adopted as illustrated in Fig. 1(d), where the actual channel length is partly correlated with the unscaled gate length, thus not being *strictly downscaled*. It should be highlighted that the suppression of non-ideality factors could be eventually achieved with numerous existing strategies; it is then the *intrinsic mobility* that would be the limitation. However, the *intrinsic carrier mobility* after *strict downscaling* has not been properly investigated.

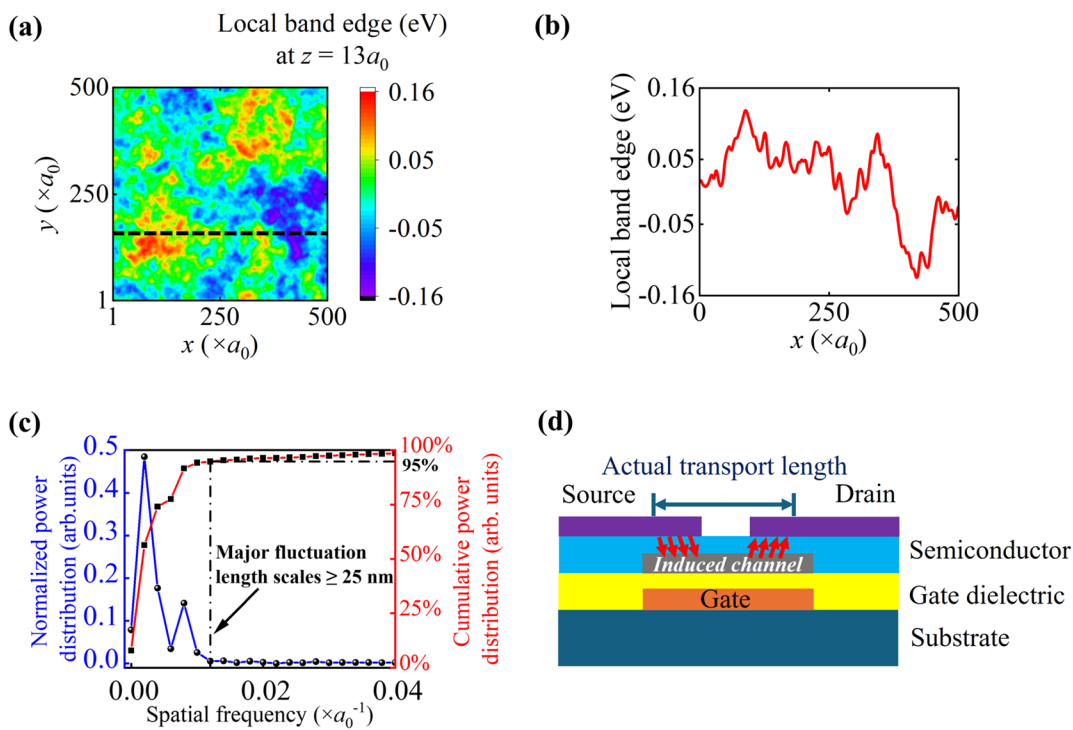

FIG. 1. Preliminary justification of mobility increase through downscaling. (a), (b) The band edge spatial distribution of the model developed in Paper 1 [25]. The $150 \times 150 \times 7.5$ nm model simulates a hydrogenated amorphous silicon ($a$-Si:H) thin film; shown in (a) is the result of the plane at $z = 13a_0$, where $a_0 = 3$ Å is the short-range order length. The one-dimensional fluctuation along the dashed line in (a) is extracted in (b). (c) Fourier decomposition of (b) showing that more than 95% contribution to the fluctuation in (b) comes from those components whose spatial frequencies are less than $0.012a_0^{-1}$. (d) Illustration of one of the reasons why many previous researchers did not observe mobility increase after downscaling. In staggered TFTs, the gate is not effectively downscaled, leading to a longer carrier path than the source-drain separation.

Therefore, a quantitative and rigorous bottom-up effort is made in this paper. First, a basic device configuration is defined in Sec. II, in the context of which the downscaling of amorphous semiconductors exemplified by *a*-Si:H is investigated. Section III resolves the motion and transport of carriers in the downscaled *a*-Si:H, followed by an analytical quantification of the intrinsic carrier mobility. Section IV uses the analytical results in Sec. III as well as the modeling framework in Paper 1 to simulate the mobility after downscaling. Remarkable mobility improvement of ~ 12 times at a length scale of 10 nm will be shown, and a satisfactorily low device-to-device variation will be revealed. In the end, it will be concluded that the essence of mobility improvement is a result of reduced localized band tail states due to the ultra-short material length scale relative to the band fluctuation length scale set by the short- and medium-range order of the amorphous semiconductor.

## II. THE CONTEXT OF DOWNSCALING – DEVICE STRUCTURE

As shown in Fig. 2(a), a basic device structure named the "nanogap" is proposed. This is a structure where a pair of metal electrodes are separated by a nanoscale length $L$ (ideally < 25 nm as envisaged earlier), in which *a*-Si:H is filled. The width ($y$ direction) and thickness ($z$ direction) are of the same scales as the source and drain electrodes in an ordinary TFT. *a*-Si:H exists only inside the gap, thus being *strictly downscaled*. To implement such strict downscaling in practical TFTs, only coplanar structures exemplified by Fig. 2(b) may be adopted. The nanoscale spacing may be fabricated using recent low-cost scalable processes such as adhesion lithography [29-32], as an alternative to the slow and expensive electron beam lithography. The gate field penetration in the TFT should be satisfactory on condition that the dielectric layer is thin and that the source and drain thickness is comparable to or less than $L$. For the purpose of this paper, we focus only on the intrinsic DC carrier mobility in Fig. 2(a) under a voltage $U$.

## III. CHARGE TRANSPORT ANALYSIS

For *a*-Si:H, electron carriers play a much more important role in conduction than hole carriers [11], and so the focus of this paper is only given to electron carriers. Given that hydrogen passivates the majority of dangling bonds in high-quality *a*-Si:H [33], in this paper we approximately treat *a*-Si:H as being *defect-free*. In the absence of defect states, the localized states in *a*-Si:H are, in essence, band tail states that originate from band fluctuation as analyzed in Paper 1 [25]. It is analyzed in another preceding work of us (henceforth named Paper 2 [34]) that, instead of being *completely confined*, electron carriers still undergo local motion even when they are in localized band tail states. Such microscopic motion is not fundamentally different from that of extended-state electrons, both of which are being scattered from disorder. Therefore, regardless of the state of the electron carriers, their paths take a zigzagged fashion due to frequent scattering, the mean free path of which is of interatomic spacing [11] ($\approx$ the short-range order length $a_0$). This is illustrated in Fig. 3(a). Such random thermal motion acts as noise that does not contribute to net conduction. An imagined "drift center" is thus introduced in Fig. 3(b) to describe the net drift motion of every electron carrier under an applied field. Acting as a reference point, the drift center of an electron carrier is defined in such a way that it is exactly at the location of the actual carrier when the *x* coordinate of the actual carrier first reaches the middle of the nanogap (*i.e.*, $x = x_0$); this is illustrated in Fig. 3(b). Except the case of $x = x_0$, the actual location of the carrier is uncertainly distributed around the drift center.

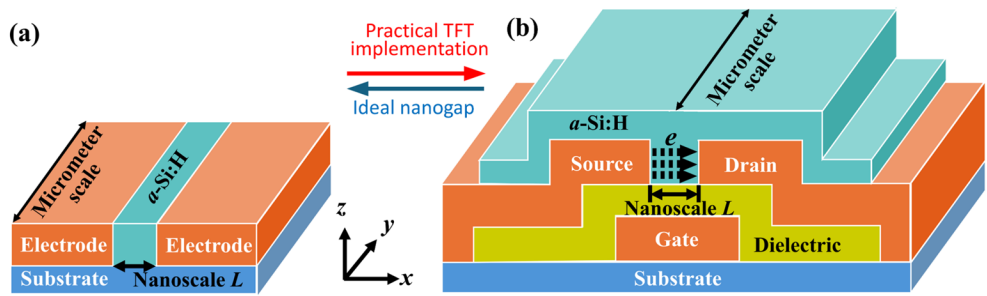

FIG. 2. Device structure and practical implementation. (a) The structure of nanogap where hydrogenated amorphous silicon (*a*-Si:H) is strictly downscaled. The length of the nanogap is in the nanoscale whereas the width and thickness are of standard scales as in ordinary thin film transistors (TFTs). (b) Practical implementation of nanogap in TFTs. A coplanar structure must be used to produce a channel that is strictly downscaled (illustrated by the dashed arrows).

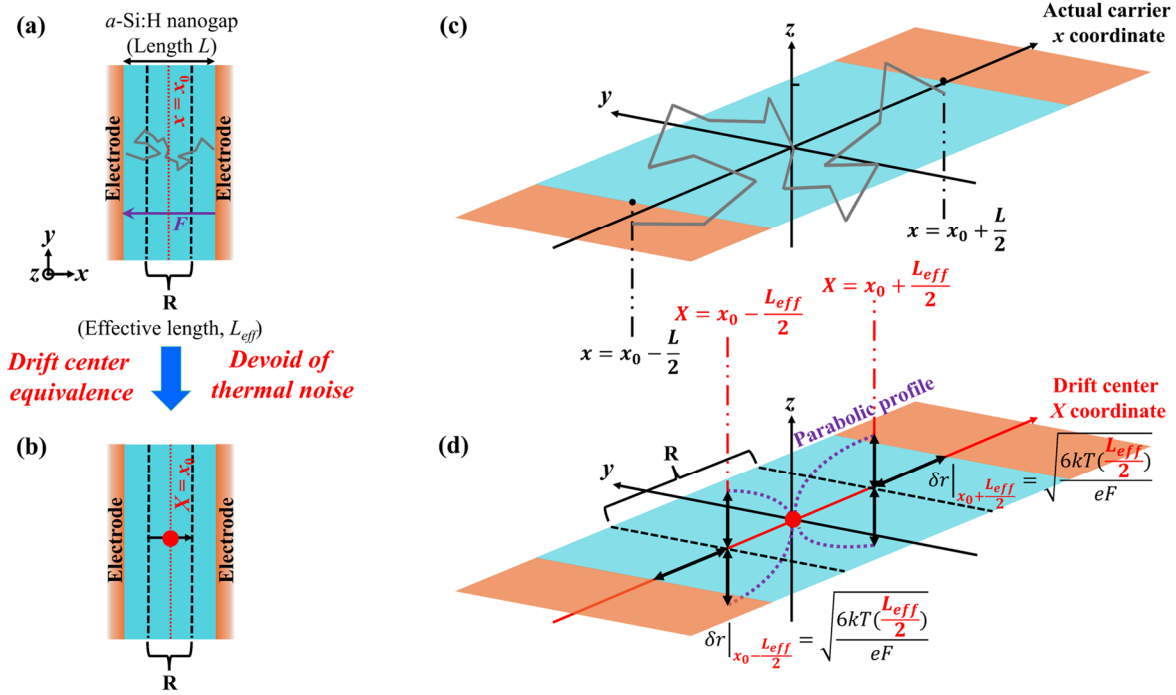

FIG. 3. Analysis of the random thermal motion and drift motion of electron carriers in the nearly defect-free hydrogenated amorphous silicon (*a*-Si:H) nanogap. (a) Illustration of the actual zigzagged path of a carrier under an applied field *F*. (b) Description of net drift motion based on an idealized drift center. The drift center (red sphere) is defined such that it is at the exact location of the actual carrier when the latter first reaches the middle of the gap (*i.e.*, $X = x = x_0$). (c) A three-dimensional (3D) illustration of the actual carrier path. During the period when the drift center drifts in a region named **R** [see also (a) and (b)], the carrier is narrowly able to reach the metal-semiconductor interfaces. (d) Quantification of the uncertainty of carrier location. Before and after the drift center reaches $X = x_0$, the root-mean-square (RMS) deviation ($\delta r$) of the actual carrier from the drift center depends on the location of drift center in a parabolic fashion (the dotted purple curves). Region **R** is defined in such a way that when the drift center reaches either of the **R** boundaries, the farthest RMS deviation is just able to reach the metal-semiconductor interfaces.

To quantitatively analyze this uncertainty, the classic random walk theorem is utilized. As revealed in Paper 2 [34], electrons at different energies experience different extents of spatial confinement. Thus, if the transit of a carrier through the *a*-Si:H nanogap includes more than one energy state, such a transit should be modeled as a random walk with multiple phases, each of which lasts for a time $t_m$ and possesses a unique diffusion coefficient $D_m$ reflecting the extent of spatial confinement. Based on Einstein's diffusion theory [35] and probability theories, the root-mean-square (RMS) displacement of carrier from its initial location in three-dimensional (3D) space is

$$\delta r = \sqrt{\sum_m 6 D_m t_m} \ . \tag{1}$$

Under an applied electric field ($F$), the diffusion coefficient $D_m$ is related to the drift mobility ($\mu_{dm}$) of the carrier during the $m^{th}$ phase via the Einstein relation

$$D_m = \frac{kT}{e} \mu_{dm} \ , \tag{2}$$

where $k$ is the Boltzmann constant, $T$ denotes temperature, and $e$ is the elementary charge. Further, the drift mobility is associated with the drift velocity ($v_{dm}$) via

$$v_{dm} = \mu_{dm} F \ . \tag{3}$$

In total, the carrier completes a *net drift* along the negative field direction by a distance $X$ that is

$$X = \sum_m v_{dm} t_m \ . \tag{4}$$

Substituting Eqs. (2), (3), (4) back to Eq. (1), the total RMS displacement is expressed as

$$\delta r = \sqrt{\frac{6kTX}{eF}} \, , \qquad (5)$$

which relates the net drift with the superimposed diffusion. This is illustrated by the dashed sphere and the dotted evolution contour in Fig. 3(d) in a 3D view [to be viewed alongside Fig. 3(c)]. The period before the drift center reaches $X = x_0$ is simply treated as a reverse process, and so the location uncertainty is similarly quantified, showing a mirrored RMS profile about $X = x_0$. It should be noted that the symmetry shown in Fig. 3(d) results from how we define the location of drift center initially. Although in principle different definitions would not change the result of this paper, the adopted definition here provides the best convenience as will be manifested soon.

It is now apparent that electron carriers may be in the metal electrodes when their drift centers are very near the metal-semiconductor interfaces. This will introduce complexity to our subsequent analysis on charge transport, because the electric field in the metal is noticeably weaker than that in $a$-Si:H; the intrinsic property of $a$-Si:H will become vague if the effect of metal is included. In fact, the latter should be recognized as a parasitic factor in the form of, $e.g.$, contact resistance. To avoid such complexity and reveal the $intrinsic$ $property$, a region named $\mathbf{R}$ with a length of $L_{eff}$ within $a$-Si:H is defined. The expectations are that, first, $\mathbf{R}$ needs to be kept away from the electrodes as illustrated in Fig. 3(a) and (b); in this way, as long as the drift center is within $\mathbf{R}$, the actual carrier should not be in the metals $in$ $most$ $cases$. Second, $\mathbf{R}$ should be as large as possible such that the intrinsic property of the entire $a$-Si:H is revealed, instead of a small portion of it. These requirements are analyzed in Fig. 3(d), where the largest RMS deviation spheres (when the drift center is on either side of $\mathbf{R}$) is exactly tangential to the metal-semiconductor interfaces. Consequently, the length of $\mathbf{R}$ can be analytically determined via

$$\frac{L - L_{eff}}{2} = \sqrt{\frac{3kT}{eF} L_{eff}} \, . \qquad (6)$$

The mobility of the entire $a$-Si:H in the nanogap is now to be analyzed based on the motion of drift centers across $\mathbf{R}$. It is assumed that, after applying the voltage $U$, the DC conduction lasts for a sufficiently long time $T_0$ relative to the time $t_0$ for the system to reach a steady current $I_{SS}$. The average conductivity ($\bar{\sigma}$) of $a$-Si:H in the nanogap is thus expressed as

$$\bar{\sigma} = \frac{I_{SS} L}{US} \, , \qquad (7)$$

where $S$ is the cross section area of the nanogap. It should be noted that the measured $I_{SS}$ is deemed as the $net$ $current$ that is contributed only by the drift motion

of electrons and is exclusive of noise caused by random thermal motion; it can therefore be calculated as

$$I_{SS} = \frac{eN_d}{T_0} \, , \qquad (8)$$

assuming that there are totally $N_d$ drift centers that passed through $\mathbf{R}$ within $T_0$.

Excluding thermal noise and statistical fluctuation, the average number of drift centers inside $\mathbf{R}$, $\bar{N}_{\mathbf{R}}$, is

$$\begin{cases} \bar{N}_{\mathbf{R}} = L_{eff} S \bar{n}_{\mathbf{R}} \\ \bar{n}_{\mathbf{R}} = \int_{E_{F,ng}}^{+\infty} g_{\mathbf{R}}(E) f(E) dE \end{cases} , \qquad (9)$$

where $g_{\mathbf{R}}(E)$ is the DOS distribution of the $a$-Si:H inside $\mathbf{R}$, $E_{F,ng}$ is the Fermi level of the $a$-Si:H in the nanogap. Note that $\bar{N}_{\mathbf{R}}$ is far less than unity considering the nanoscale $L_{eff}$ and the limited DOS. Therefore, the actual number of drift centers within $\mathbf{R}$ is temporally varying between 0 and 1. Suppose that the drift center of the $i$th of the $N_d$ electrons drifts across $\mathbf{R}$ for a duration $t_i$, where

$$t_i = \frac{L_{eff}}{\mu_i F} = \frac{L_{eff} L}{\mu_i U} \, , \qquad (10)$$

with $i \in [1, N_d]$ and $\mu_i$ being the unique mobility of the drift center as it crosses $\mathbf{R}$. Note that although the drift center only exists in $\mathbf{R}$ in the analysis here, the corresponding actual electron carrier is able to span wider within the entire nanogap. Consequently, the average electric field experienced by the actual electron is the total electric field $F$ across the entire nanogap.

An interpretation of the statistical average $\bar{N}_{\mathbf{R}}$ from a temporal perspective shows that the cumulation of the drift time of all the $N_d$ electrons is equal to $\bar{N}_{\mathbf{R}} T_0$. Thus,

$$\sum_{i=1}^{N_d} t_i = \bar{N}_{\mathbf{R}} T_0 \, . \qquad (11)$$

Substituting Eqs. (8) and (10) into Eq. (11) yields that

$$\sum_{i=1}^{N_d} \frac{L_{eff} L}{\mu_i U} = \bar{N}_{\mathbf{R}} \frac{eN_d}{I_{SS}} \, . \qquad (12)$$

The average electron mobility of $a$-Si:H in the entire nanogap is defined by Eq. (7) as

$$\bar{\mu} = \frac{I_{SS} L}{US} / e\bar{n} \, , \qquad (13)$$

where $\bar{n}$ represents the average density of electron carriers in the entire nanogap, which is

$$\bar{n} = \int_{E_{F,ng}}^{+\infty} g(E) f(E) dE \, , \qquad (14)$$

with $g(E)$ being the DOS of the entire nanogap [may be slightly different from $g_{\mathbf{R}}(E)$]. Integrating Eqs. (9) and (13) into Eq. (12) leads to

$$\bar{\mu} = \frac{\bar{n}_{\mathbf{R}}}{\bar{n}} \frac{N_d}{\sum_{i=1}^{N_d} \frac{1}{\mu_i}} . \qquad (15)$$

Multiple trapping and release (MTR) is the predominant charge transport mechanism of room-temperature *a*-Si:H [36] (and many other amorphous semiconductors). According to Paper 2, the average DC mobility of electrons in a *macroscale a*-Si:H thin film is [34]

$$\mu_a = \mu_{Ceff} \frac{n_{ext,eff}}{n_{ext,eff} + n_{loc,eff}} , \qquad (16)$$

where $\mu_{Ceff}$ is the effective extended-state mobility. $n_{ext,eff}$ and $n_{loc,eff}$ respectively denote the densities of electrons that are in effective extended states and effective localized states, demarcated by the effective conduction band (CB) mobility edge $E_{Ceff}$; these densities are calculated based on the inner product of DOS and Fermi distribution.

It is tempting to evaluate the mobilities $\mu_i$ in Eq. (15) using Eq. (16). However, Paper 2 analyzed that Eq. (16) is valid only when both of the following assumptions are true [34]. First, uniformity. Extended states and localized states are approximately uniform in space such that all carriers end up interacting with generally similar electronic structures. Second, sufficiency. Carriers undergo numerous trapping and release events during their migration to allow the exhibition of statistical nature. However, these assumptions collapse in the downscaled nanogap. It is most evident that the sufficiency assumption collapses as an electron is unlikely to undergo many trapping and release events during its ultra-short migration time across the gap. The uniformity assumption also no longer holds true. The dimension along the *x* axis is too small to leverage any macroscale uniformity. Though there is macroscopic uniformity along the *y* axis, the diffusion of electron along this direction is limited given the analysis in Fig. 3(d). Thus, electrons that are well-separated in the *y* axis may end up interacting with noticeably different electronic structures.

To solve the issue of nonuniformity, in Fig. 4(a), the nanogap is *virtually* divided along the *y* and *z* axes into $N_\perp$ "lanes" with a reasonable cross section. The expectation is that, in this way, carriers can be treated as being *confined* within the respective lanes along which their drift centers migrate; uniformity is achieved inside the individual lanes. To meet this expectation, the lane cross section should be the *average* transverse deviation of carriers away from their drift centers. According to Eq. (5) and Fig. 3(d), this is defined based on the volume average, *i.e.*,

$$\frac{4}{3}\pi\left\langle\left(\delta r\right)^3\right\rangle = \frac{\int_{x_0 - \frac{L_{eff}}{2}}^{x_0 + \frac{L_{eff}}{2}} \frac{4}{3}\pi\left(\sqrt{\frac{6kT|X - x_0|}{eF}}\right)^3 dX}{L_{eff}} \qquad (17)$$

through which the cross-sectional side length of the lanes is set as

$$d_{cs} = 2\sqrt[3]{\left\langle \left(\delta r\right)^3\right\rangle} = \sqrt[3]{\frac{16}{5}}\sqrt{\frac{3kT}{eF}L_{eff}} \ . \qquad (18)$$

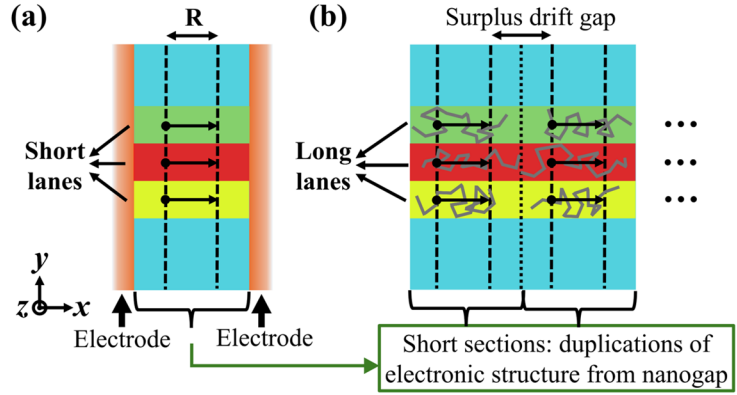

FIG. 4. Discretized transport analysis and its equivalence. (a) Division of lanes in the nanogap (*z*-axis division not shown). Each lane is given a specific color to demonstrate its unique local electronic structure. The motion of the drift centers of carriers are illustrated by the black arrows. (b) A virtual thin film constructed through duplicating the electronic structures of *a*-Si:H in (a) for numerous times along the *x* axis. The actual carrier motions are denoted by the zigzagged grey lines. The repeated drifts in (a) are equivalently reproduced in (b) by the discontinuous drifts along the respective long lanes. A continuous drift in (b) also includes drift processes in the surplus gaps which are similar to the discontinuous drifts if the length of **R** is around half of the nanogap length.

The issue of insufficiency is now to be addressed. Assume that throughout the DC conduction time $T_0$ there are totally $n_s$ drift centers that migrate across $\mathbf{R}$ along the $s^{\text{th}}$ lane, the $j^{\text{th}}$ of which exists in $\mathbf{R}$ for a duration $t_{s,j}$ and exhibits a mobility of $\mu_{s,j}$ with $j \in [1, n_s]$. Note that $n_s$ should be sufficiently large in the time $T_0$, so the statistical nature

$$\sum_{j=1}^{n_s} t_{s,j} = \bar{N}_{\mathbf{R},s}T_0 \qquad (19)$$

can be exhibited, where $\bar{N}_{\mathbf{R},s}$ is the average number of drift centers inside the $s^{\text{th}}$ lane within $\mathbf{R}$, which is quantified by

$$\bar{N}_{\mathbf{R},s} = \frac{L_{eff}S}{N_\perp}\int_{E_{F,ng}}^{+\infty} g_{\mathbf{R},s}\left(E\right)f\left(E\right)dE \ , \qquad (20)$$

where $g_{\mathbf{R},s}(E)$ is the local DOS distribution of the $s^{\text{th}}$ lane within $\mathbf{R}$.

Neglecting the influence of thermal fluctuations in the material, there should not be more than one drift center existing in $\mathbf{R}$ at any single moment because of the limited carrier density and the extremely small volume of $\mathbf{R}$. As a result, basic statistics reveals that the migrations of the $n_s$ drift centers are *independent random experiments*, while the drift time (thus the mobilities) of these electrons are *outcomes*. As electrons are indistinguishable, it is equivalent to conduct these random experiments using a single electron; this corresponds to a scenario where a single drift center migrates across $\mathbf{R}$ along the $s^{\text{th}}$ lane for $n_s$ times.

This scenario is reproduced virtually. In Fig. 4(b), the *electronic structures* (not the real material) of $a$-Si:H in the nanogap are duplicated $M$ times along the $x$ axis, where, for $\forall s \in [1, N_\perp]$, $M > n_s$. A "virtual thin film" forms, which comprises "long lanes" duplicated from the electronic structure of the corresponding short lanes in Fig. 4(a). The scenario of "a single drift center migrating across $\mathbf{R}$ for $n_s$ times" is now equivalently reproduced in Fig. 4(b) as the discontinuous drift of a single drift center across $n_s$ short sections in a long lane under the same electric field $F$. The drift is discontinuous in the sense that the drift center is expected to instantly "jump over" the surplus gaps of length ($L - L_{eff}$), even though the actual carrier may interact with the electronic structures in these surplus gaps. Note that, because the long lanes in the virtual thin film form via the duplication of short lanes in the nanogap, the carrier will also be confined within a single long lane in the virtual thin film, just as it would be in a short lane of the nanogap. Such a limited RMS deviation over a long drift distance is different from the situation in a real macroscale thin film, which may be imaginarily understood as a result of the existence of certain "boundaries" between adjacent long lanes, which have no other effects except elastically scattering electrons.

In Fig. 4(b), the drift center requires a time $t_{s,k}$ to cross the $k^{\text{th}}$ section (of length $L_{eff}$) in the $s^{\text{th}}$ long lane, and exhibits a mobility $\mu_{s,k}$ with $k \in [1, n_s]$. Given that $n_s$ is sufficiently large, the statistical equivalence described above should be exhibited, which yields

$$\sum_{k=1}^{n_s} t_{s,k} = \sum_{j=1}^{n_s} t_{s,j} \ , \qquad (21)$$

with both equal to $\bar{N}_{\mathbf{R},s}T_0$ considering Eq. (19). Further, given the same total electric field $F$ in both Fig. 4(a) and (b), $U/L_{eff}L$ is multiplied to both sides of Eq. (21), and so

$$\sum_{k=1}^{n_s}\frac{1}{\mu_{s,k}} = \sum_{j=1}^{n_s}\frac{1}{\mu_{s,j}} \ . \qquad (22)$$

Note that the drift along the $s^{\text{th}}$ long lane has so far been considered as being discontinuous as illustrated in Fig. 4(b). Now, instead, consider an imagined scenario where the drift center does continuously drift along the $s^{\text{th}}$ long lane in Fig. 4(b); this drift center will exhibit an average mobility after it crosses $n_s$ duplicated sections. Again, because $n_s$ is sufficiently large, this average mobility will asymptotically approach a constant value termed the *lane mobility*, $\mu_s$, which reflects the unique property of the specific long lane.

Such an imagined continuous drift process includes both the discontinuous drift and the drift across the surplus gaps, and so it can be obtained that

$$\sum_{k=1}^{n_s}\frac{L_{eff}}{\frac{U}{L}\mu_{s,k}} + \sum_{k'=1}^{n_s}\frac{L - L_{eff}}{F'\mu_{s,k'}} = \frac{n_sL}{\frac{U}{L}\bar{\mu}_s} \ , \qquad (23)$$

where $k'$ indexes the drift across each of the surplus gaps, during which the drift center exhibits a unique mobility $\mu_{s,k'}$. $F'$ denotes the electric field that the carrier experiences whilst the drift center is in these surplus gaps. It will be shown later in the next subsection that $L_{eff}$ is around half of $L$, and so it is reasonably assumed that the migrations of the drift center within the surplus gaps are similar to the discontinuous drift in Fig. 4(b); this leads to the approximations

$$\begin{cases} F' = \dfrac{U}{L} \\ \displaystyle\sum_{k=1}^{n_s} \dfrac{1}{\mu_{s,k}} = \sum_{k'=1}^{n_s} \dfrac{1}{\mu_{s,k'}} \end{cases}. \tag{24}$$

These simplify Eq. (23) and lead to

$$\sum_{k=1}^{n_s} \frac{1}{\mu_{s,k}} = \frac{n_s}{\overline{\mu}_s}, \tag{25}$$

linking which to Eq. (22), Eq. (15) is equivalently

$$\overline{\mu} = \frac{\overline{n}_{\mathbf{R}}}{\overline{n}} \frac{\displaystyle\sum_{s=1}^{N_i} n_s}{\displaystyle\sum_{s=1}^{N_i} \dfrac{n_s}{\overline{\mu}_s}}. \tag{26}$$

It should be noted that for each part of the discontinuous drift,

$$t_{s,k} = \frac{L_{eff}}{\dfrac{U}{\mu_{s,k}} \overline{L}}, \tag{27}$$

and so summing up $k$ on both sides yields

$$\sum_{k=1}^{n_s} t_{s,k} = \frac{L_{eff} L}{U} \sum_{k=1}^{n_s} \frac{1}{\mu_{s,k}}. \tag{28}$$

Combining this with Eq. (25),

$$n_s = \overline{\mu}_s \frac{U}{L_{eff} L} \sum_{k=1}^{n_s} t_{s,k}. \tag{29}$$

Therefore, Eqs. (19), (21) and (29) simplify Eq. (26) to

$$\overline{\mu} = \frac{\overline{n}_{\mathbf{R}}}{\overline{n}} \frac{\displaystyle\sum_{s=1}^{N_i} \overline{N}_{\mathbf{R},s} \overline{\mu}_s}{\displaystyle\sum_{s=1}^{N_i} \overline{N}_{\mathbf{R},s}}. \tag{30}$$

Because each of the lanes in Fig. 4(b) is sufficiently long in the drift direction to allow the exhibition of long-range uniformity and the statistical nature, $\overline{\mu}_s$ can now be directly quantified via the MTR theory shown in Eq. (16). $\overline{\mu}$ can then be calculated using Eq. (30).

## IV. RESULTS AND DISCUSSIONS

### A. Simulation of a 10-nm nanogap

As a comparative study, we base this paper on the 2.5-dimensional (2.5D) excess delocalized charge model developed in Paper 1 [25]. This model approximates a defect-free $a$-Si:H thin film, whose electron mobility is to be compared with that after downscaling in the nanogap. Note that due to computational cost, the length scale of the 2.5D model was only 150 nm (rather than micrometer scale in real situations). Hence, to produce a significant contrast, here a 10 nm nanogap is first simulated, though in practice a slightly longer gap enhances gate controllability. It should be mentioned that the $a$-Si:H in the nanogap is expected to be fabricated through etching from a large thin film, and so the extent of bond angle and bond length distortion, which determines the excess delocalized charges, is the same in both cases. Thus, $a$-Si:H in the nanogap is simply modeled as a "strip" cut from the 2.5D excess delocalized charge model [Fig. 5(a)], assuming trivial metal-semiconductor interfacial charges.

As the first trial, we extract the strip between $x = 50a_0$ and $x = 83a_0$ from the 2.5D model. The band edge distribution for the $z = 13a_0$ plane is mapped in Fig. 5(b), based on which $g_{\mathbf{R}}(E)$ and $g(E)$ can be calculated using the method in Paper 1 [25]. Lane division is specified according to Eq. (18) assuming that $U = 0.5$ V, based on which $g_{\mathbf{R},s}(E)$ is obtained. For computational simplicity, the energy integrations in Eqs. (9), (14) and (20) start from the energy position at which the corresponding DOS is equal to $10^{18}$ eV$^{-1}$cm$^{-3}$. This is reasonable because for a nearly defect-free $a$-Si:H, DOS below $10^{18}$ eV$^{-1}$cm$^{-3}$ continues to drop exponentially and is thus much more trivial compared with the DOS at higher energies; for example, the latter is typically $> 10^{21}$ eV$^{-1}$cm$^{-3}$ at extended-state energies [25]. Note that although the electric field at $U = 0.5$ V (i.e., 0.5 MV/cm) is high enough to cause the onset of velocity saturation of extended-state electrons in $a$-Si:H due to acoustic phonon emission, the effect of this saturation is not strong as shown in Ref. [37]; meanwhile, a high field is able to promote thermal release, which overcompensates the velocity saturation effect. Therefore, if combining these two counteracting factors into the single quantity $\mu c_{eff}$ in Eq. (16), this quantity should be higher under stronger field. To avoid causing extra complexity, in this paper we treat $\mu c_{eff}$ as a field-independent quantity; if, even under this assumption, the intrinsic mobility $\overline{\mu}$ is revealed to increase after downscaling, the actual case should be even more promising. Furthermore, $\mu c_{eff}$, being a quantity related to carrier scattering, may also be treated as being dimension-independent here, because the mean free path of free electrons in amorphous semiconductors is of interatomic spacing scale, still being much shorter than the gap length (i.e., there is no ballistic transport despite the device downscaling). Hence, $\mu c_{eff}$ will be canceled if we focus on the relative mobility increase. Still, we continue to use the value of 46 cm$^2$/(V s) as revealed in Paper 2 [34], in order to straightforwardly show the absolute mobilities.

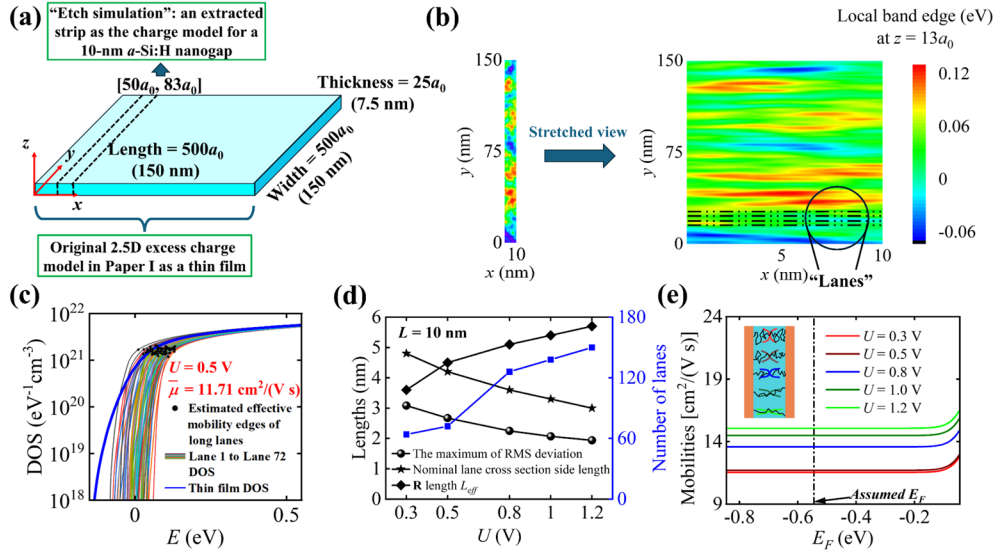

FIG. 5. Quantification of nanogap mobility. (a) Modeling *a*-Si:H nanogap based on the 2.5-dimensional (2.5D) excess charge model in Paper 1 [25]. (b) The band edge distribution of *a*-Si:H in the nanogap. The distribution on the middle plane ($z = 13a_0$) is shown in both the original and stretched aspect ratio. The division of lanes is illustrated in the latter. (c) Calculated density of states (DOS) distributions of individual lanes under 0.5 V. These DOS curves are much steeper compared with that of the thin film counterpart. Overestimated mobility edges (black dots) are used to underestimate the 72 lane mobilities. (d) Investigation of the effect of voltage. A higher voltage confines the deviation of carriers from their drift centers, shrinks the lane cross sections (thus increasing the number of lanes), and increases the length of **R**; these are illustrated in the inset of (e). (e) Dependence of nanogap mobility on voltage and Fermi level position. As is envisaged based on (d), a higher voltage increases nanogap mobility. The Fermi level position (within a reasonable range) has almost zero influence on nanogap mobility. The energies are relative to the conduction band edge of crystalline silicon. The arrow indicates the choice of Fermi level in this paper. The parabolic curves in the inset show the evolution of RMS deviation under different voltages.

Fig. 5(c) shows $g_{\mathbf{R},s}(E)$, the local DOS's of individual lanes, which are compared with the DOS of the thin film counterpart. It is evident that downscaling fundamentally changes the steepness of these constituent DOS's, which validates our hypothesis proposed in the beginning that high-magnitude low-frequency band fluctuation components are filtered through downscaling. To evaluate room-temperature lane mobilities $\overline{\mu_s}$ using Eq. (16), the effective mobility edge ($E_{Ceff}$) of the lanes needs to be known in addition to $\mu_{Ceff}$. A worst-case scenario is considered. $E_{Ceff}$ is overestimated to be at the highest local band edge in the individual lanes; this is shown by the black dots on the DOS curves in Fig. 5(c). Accordingly, the nanogap intrinsic mobility $\overline{\mu}$ is revealed to be 11.71 cm²/(V s) through Eq. (30), *i.e.*, ~ 12 times that of a large-scale thin film.

### B. Simulation of device-to-device (D2D) variation

"Cutting" from different parts of the 2.5D model is analogous to batch-fabricating nanogaps on a single substrate. The D2D variation of the intrinsic mobility is thus simulated here through extracting charges at more locations, for instance, at $x \in [100a_0, 133a_0]$, $[200a_0, 233a_0]$, $[290a_0, 323a_0]$, $[350a_0, 383a_0]$ and

$[420a_0, 453a_0]$. This is illustrated in Fig. 6(a). Together with the first trial and under the same $U = 0.5$ V, six nanogap mobilities are shown in Fig. 6(b); the mobility of the thin film counterpart as well as $\mu_{Ceff}$ are indicated as comparisons. On average, these nanogaps exhibit an average mobility of 11.86 cm²/(V s), with the highest (lowest) mobility being ∼ 14 times (9 times) that of the thin film counterpart ($\mu_{tf}$). The coefficient of variation (CV), which quantifies the material-level D2D, is 16.26%. To quantitatively make sense of this CV value, the variation of the 72 lane mobilities ($\overline{\mu_s}$) within each single trial is investigated as a comparison. These lane mobility values are indicated by the scatters in Fig. 6(b); their CV values are recorded in the top table. Compared with these, the CV of nanogaps is significantly smaller; this results from the sufficient width of the nanogaps. Adventitious mobility variations between lanes are likely to be suppressed through the weighted mean operation in Eq. (30), considering that the correlation between $\overline{\mu_s}$ and their weights in Eq. (30) is weak as shown in Fig. 6(c). Thus, it is expected that for real nanogaps with much larger widths (*e.g.*, in the micrometer scale) than the simulation in this paper, the real D2D variation should be sufficiently low.

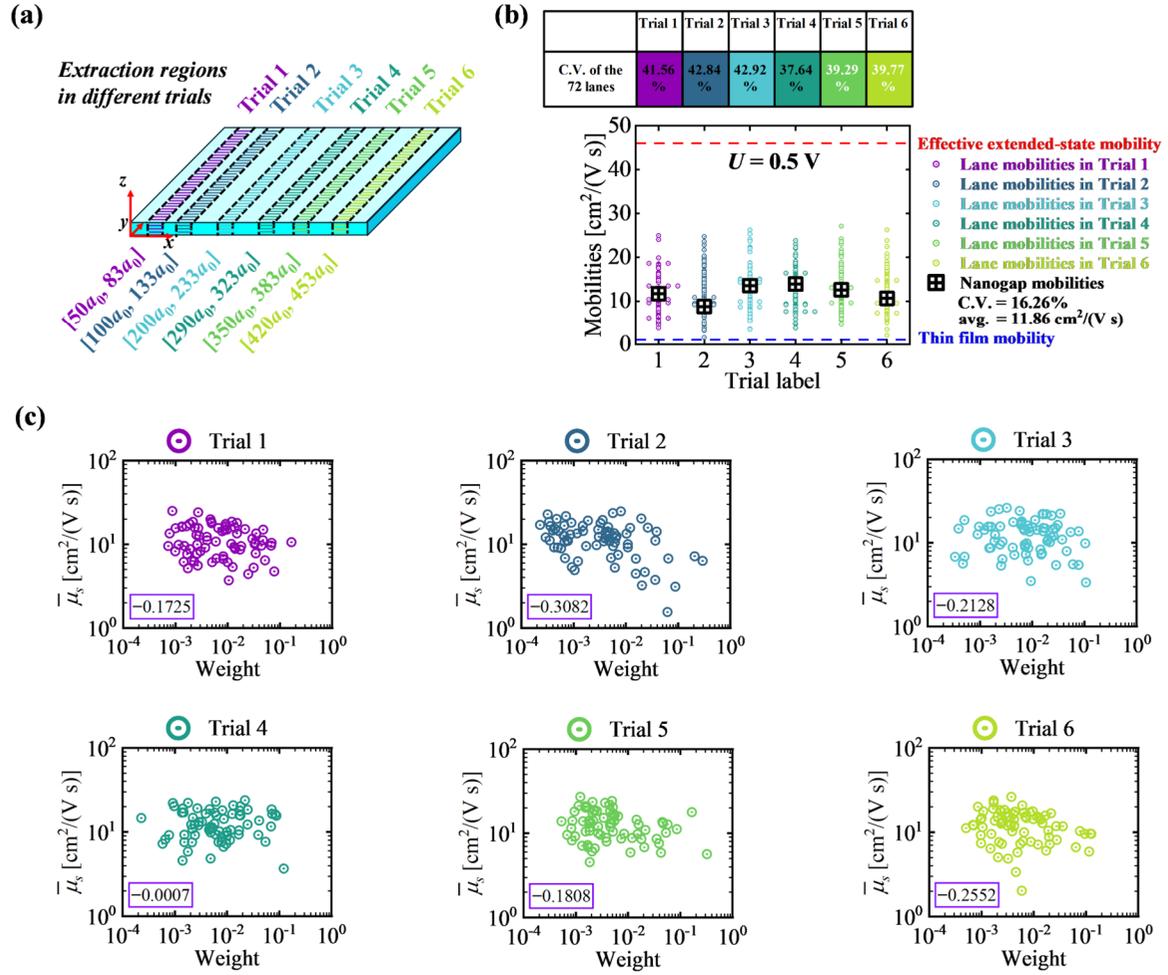

FIG. 6. Study of device-to-device (D2D) variation at the material level. (a) Illustration of the production of six random trials. (b) Mobility variation. The average mobility of the six trials is 11.97 times the mobility ($\mu_{tf}$) of their thin film counterpart (dashed blue line). The coefficient of variation (CV), which reflects the D2D variation, is 16.26%. The individual lane mobilities in each trial are denoted by the hollow circles. The top table lists the CV of the lane mobilities for each trial; 16.26% is significantly low in comparison. (c) Correlation between the lane mobilities ($\bar{\mu}_s$) and their corresponding weights in eq 30. All the six random trials are investigated. The correlation is weak, as revealed by the Pearson coefficients labeled in the frames. Logarithmic scales are adopted for a better presentation.

## C. Dependence of D2D variation on the nanogap width

To better demonstrate the influence of nanogap width on the D2D variation, nanogaps with five more different widths are simulated. The way that these nanogaps are produced is illustrated in Fig. 7(a). Fig. 7(b) shows that the mobility CV of the six nanogaps is low either in the case of a sufficiently large width (as expected earlier), or in the case of a sufficiently small width. The mechanisms of such a trend are explained in Fig. 7(c) and (d). Fig. 7(c) shows the band fluctuation of an example nanogap, from which deep energy valleys are identified. If a lane of a nanogap contains one such electronic structure, the lane mobility would be considerably low, because a valley produces significant localized states which retard the drift of carriers. Further, the $\bar{N}_{R,s}$ of this lane would be high compared with other lanes in the nanogap, because the local DOS distribution of this lane is closer to the Fermi level, indicative of a higher occupation probability $f$. Thus, the mobility of the entire nanogap would be nontrivially affected by the presence of deep energy valleys; different nanogaps behave differently if some does not contain a deep valley while others contain one or more.

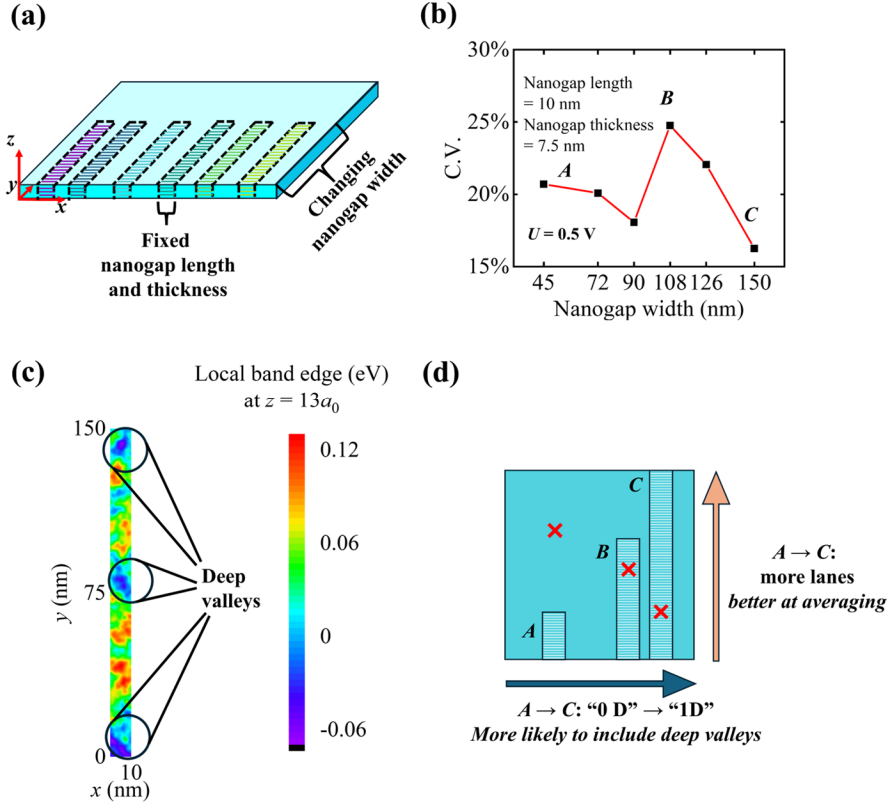

FIG. 7. Dependence of device-to-device (D2D) variation on nanogap width. (a) Illustration of the production of nanogaps with different widths. (b) Coefficient of variation (CV) calculated at different widths. The symbols *A*, *B* and *C* denote three regimes featuring small, medium and large widths. (c) Examples of deep energy valleys in the band fluctuation landscape. (d) Illustration of the existence of deep valleys (red "×" symbols) in a thin film and the mechanisms causing or suppressing D2D variation. The three regimes mentioned in (b) are illustrated, represented by extracted regions with different widths. On the one hand, a small width is better at avoiding deep valleys which cause property discrepancies; on the other hand, a large width is better at suppressing discrepancies between lanes through the weighted mean operation in Eq (30).

As illustrated in Fig. 7(d), on the one hand, there is a higher chance for a nanogap with a larger width to contain deep valley(s) if viewing the valleys as randomly scattered points. On the other hand, a wider nanogap contains more lanes, such that through the weighted mean operation in Eq. (30), the property fluctuation caused by a single (or a few) lane(s) would be suppressed. These two mechanisms counteract each other. For a considerably small width, the former mechanism wins: it is less likely to produce a nanogap which contains a deep valley. By contrast, for a considerably large width, the latter mechanism wins. Although a low D2D variation may be achieved in either way, utilizing a large width is favored from the perspective of practical applications, because of the easier fabrication compared with downscaling in all dimensions.

### D. Dependence of mobility on the nanogap length

Five more length scales, 20 nm, 30 nm, 40 nm, 50 nm and 60 nm, are studied. In order to obtain statistically reliable results without being affected by stochastic variations, six nanogaps are simulated for each of these lengths. Because the 2.5D model is limited in size due to computational load, also in light of the space occupations of the extraction regions in the 2.5D model, the width of the six nanogaps for each length decreases to only 45 nm in the investigation of this subsection; they are extracted according to the illustration in Fig. 8(a). Together with the results of the 10-nm nanogaps, totally 6×6 nanogap mobilities under the same voltage $U = 0.5$ V are calculated in Fig. 8(b). The average mobility of the six nanogaps at each length is then calculated, fitting well with an exponential function of the form (dashed blue curve)

$$\bar{\mu} = \mu_{const.} \exp\left(-\frac{L}{L_0}\right) + \mu_{if} \, , \qquad (31)$$

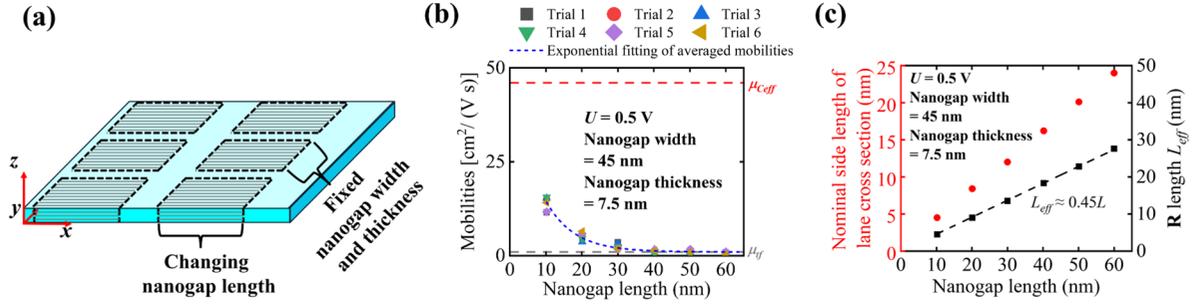

FIG. 8. Dependence of nanogap mobility on nanogap length. (a) Illustration of the production of nanogaps with different lengths. Due to the limit of simulation size, each nanogap is only 45nm-wide, such that six trials can be produced from a single thin film model. (b) Simulation results and fitting. For each of the six nanogap lengths, six trials are conducted. Totally 36 mobility results are obtained. The average mobility of the six trials at each of the six lengths is calculated and fitted using an exponential decay function (eq 31). The mobility of the thin film counterpart ($\mu_{tf}$) and the effective extended-state mobility ($\mu_{C\,eff}$) are shown as comparisons. (c) Dependence of the length of **R** and the side length of lane cross section on the length of nanogap. Both of them scales linearly with the nanogap length under a fixed voltage $U = 0.5$ V.

where $L_0$ is a length constant and $\mu_{const.}$ is a mobility constant. Note that extrapolation of lengths below 10 nm is not shown because these lengths are too short to be practically meaningful. The above results further validate our hypothesis that downscaling acts as a spatial frequency filter. The greater the downscaling, the fewer the low-frequency components, the lower the fluctuation amplitudes, and the higher the mobility.

It is worthwhile to note that the size of 20 nm (length) × 45 nm (width) × 7.5 nm (thickness) is simulated in this subsection, showing an average mobility of 5.16 cm$^2$/(V s). In comparison, the size of 10 nm (length) × 90 nm (width) × 7.5 nm (thickness) was simulated in the last subsection [Fig. (7)], which shares the same volume as the former case, but which exhibits a much higher mobility of 12.27 cm$^2$/(V s). The fundamental reason why the same volume corresponds to different mobilities lies in the *directionality* defined by the carrier drift direction. In order to increase mobility, it is important to note which dimension is to be downscaled; this must be in the direction along which carriers are drifting (*i.e.*, opposite to the direction of the applied electric field).

Shown in Fig. 8(c) is a demonstration that, under the same $U$, the length of **R** scales linearly as the nanogap length increases, and that the nominal lane cross section length also scales linearly. Here, the term "nominal" is used because the discretized division could result in a few "incomplete" lanes on the margins, but these are minorities under most circumstances.

### E. The role of medium-range order (MRO)

According to Paper 1, the medium-range order (MRO) of *a*-Si:H may be quantitatively evaluated using a modeling parameter named *window dimension* (*w*) [25]. Viewing this parameter as a variable that is related to the conditions and quality of material deposition, the dependence of nanogap mobility on *w* is investigated in this subsection. Here, in addition to $w = 5a_0$ which was discussed in Paper 1 and in previous subsections of this paper, another four choices of *w* (*i.e.*, $w = 0$, $2a_0$, $10a_0$ and $20a_0$) are studied. Firstly, 2.5D thin film models based on these *w*'s are established following a principle that these thin films all possess the same DOS distribution as the $w = 5a_0$ case did, thus exhibiting the same mobility at the macroscale. This is for the purpose of exclusively revealing the influence of MRO. Then, the difference in MRO is observed from the band fluctuation landscapes [Fig. 9(a), (b), (c), (d) and (e), before downscaling]. It is evident that a larger *w* (indicative of a longer MRO length scale) corresponds to a smoother band fluctuation. From the perspective of Fourier decomposition and as analyzed in Fig. 9(f) and (g), this means that, with a larger *w*, high-frequency components are more trivial. It is thus envisaged that after downscaling, these remnant high-frequency components will possess lower amplitudes; this will lead to steeper lane DOS's and, consequently, a higher nanogap mobility.

The above envisagement is confirmed by the results in Fig. 10. In Fig. 10, it is shown that, at any fixed nanogap length $L$, the mobility increases nearly linearly with the increase of *w*. It can thus be concluded that, fundamentally, it is the *relative length scale* of the nanogap compared with the MRO length, that determines the mobility improvement after downscaling.

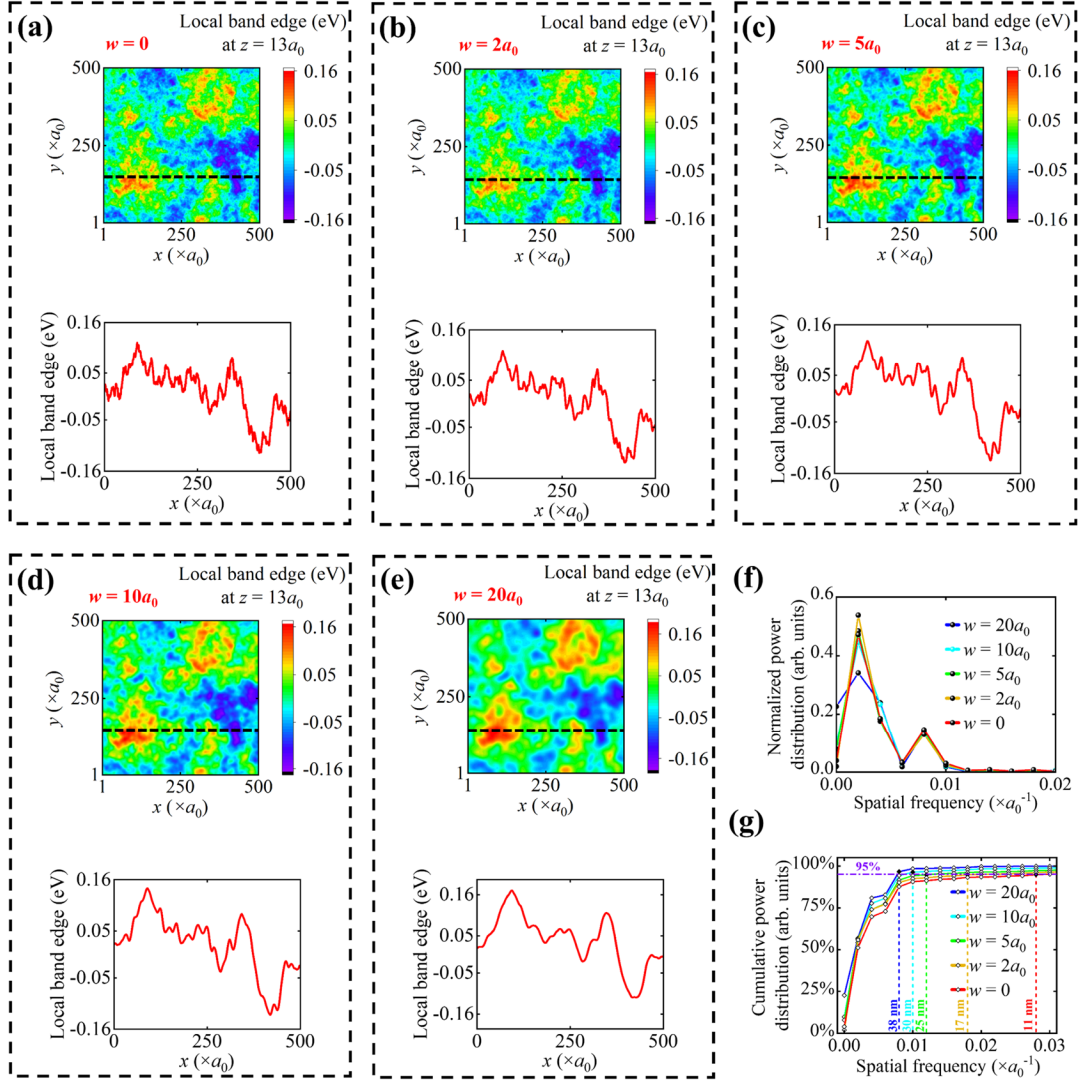

FIG. 9. Influence of medium-range order (MRO) on the spatial frequencies of band fluctuation. (a) – (e) The two-dimensional (2D) and one-dimensional (1D) band fluctuation of thin film models that are produced based on different window dimension ($w$). (f) Fourier decomposition of the 1D fluctuations in (a) – (e). (g) Cumulative distribution of spatial frequencies in (f). The spatial frequency at which the cumulation reaches 95% is defined as the cutoff frequency. As $w$ increases, the cutoff frequency decreases, and the length scale which corresponds to the cutoff frequency increases as labeled in the figure.

## F. Discussions

From the above analyses and results, it is now clear that downscaling improves the intrinsic carrier mobility of an amorphous semiconductor because of two reasons. First, the nanoscale length leads to an ultrashort migration time of carriers across the gap; unlike in a macroscale film where the lateral diffusion of carriers can be noticeable, the carriers in the nanogap predominantly move along the downscaled length direction. Second, along this length direction, the band fluctuation of the amorphous semiconductor is smoother after downscaling, resulting in a lower density of localized band tail states and consequentially enabling a higher carrier mobility. This mitigated fluctuation stems from the size shrinking which is analogous to a high pass filter that selectively blocks low-frequency fluctuation components that otherwise contribute to most of the band fluctuation. Ultimately, the extent of mobility increase depends on the detailed spatial frequency compositions of band fluctuation before downscaling and how their corresponding length scales compare to the downscaled material length; this is where the medium-range order of the amorphous semiconductor plays a role in part.

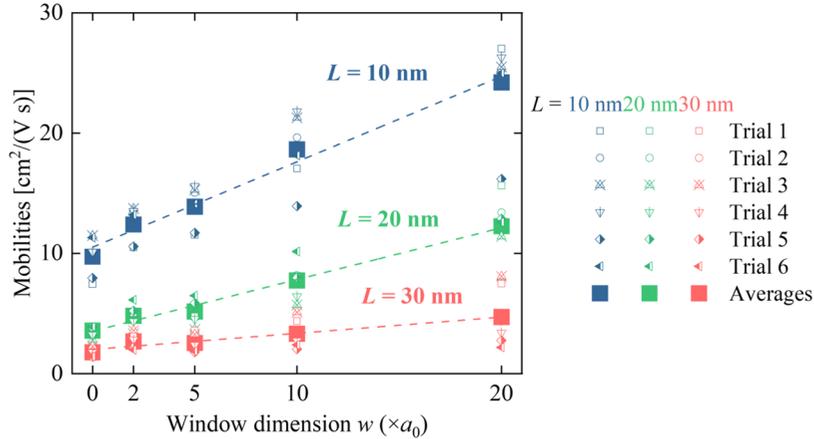

FIG. 10. Dependence of nanogap mobility on the nanogap length and the window dimension ($w$). Three nanogap lengths and five $w$'s are studied. For each case, six trials are conducted as in Figure 8. For a fixed nanogap length, the average nanogap mobility of the respective six trials increases nearly linearly with an increasing $w$. The mobility is ultimately a combined result of nanogap length and the length scale of medium-range order which is associated with $w$.

One of the superiorities of amorphous semiconductors to other non-crystalline and crystalline semiconductors is their excellent large-area uniformity, but little was known when it comes to the nanoscale. It is shown in this paper that, in order to still retain the uniformity even after downscaling, it is essential and practically applicable that at least one dimension is unscaled.

## V. CONCLUSIONS

This paper is a significant theoretical endeavor that elucidates the improvement of intrinsic carrier mobility in downscaled amorphous semiconductors. To start with, the concept of spatial frequency filtering is introduced, which proposes that the band fluctuation (which determines carrier mobility) can be decomposed into various Fourier components, and that downscaling acts as a "high pass filter" which leaves only high-frequency components with limited amplitudes that indicate a higher mobility. This preliminary hypothesis is validated in the context of a nanogap structure, using the model of $a$-Si:H established in Paper 1 [25]. It is shown that the intrinsic electron mobility of $a$-Si:H in a 10-nm nanogap can be $\sim 12$ time that of a standard $a$-Si:H thin film. Device-to-device variation is studied which reveals the importance of maintaining at least one of the other two dimensions unscaled. Furthermore, it is analyzed that the mobility improvement depends on both the extent of downscaling and the length scale of medium-range order in the amorphous semiconductor. The results in this paper are significant in light of the need of improving the performance of electronic devices based on amorphous semiconductors. For instance, combined with further device-level optimizations, high-density, high-speed, flexible microprocessors may be developed based on this paper, which may find extensive applications in next-generation wearable electronics with edge computing functionalities. Hardware development like this is increasingly important in supporting the soaring development of IoT.

*Contact email: YL778@cam.ac.uk*

## ACKNOWLEDGMENTS

The authors appreciate Dr. Gwenhivir Wyatt-Moon and Dr. Kham Niang for their insights from experiment perspectives. This work is supported by the UKRI Engineering and Physical Sciences Research Council (EP/W009041/1), the Rank Prize Return to Research Grant, and the Cambridge Commonwealth, European and International Trust Ph.D. scholarship.

## APPENDIX: SIMULATION PROCEDURES

Fig. 5, 6, 7 and 8 continue to use the 2.5D excess delocalized charge model developed in Paper 1 [25]. The nanogap mobility results in these figures are obtained in four steps.

First, produce the excess delocalized charge model for individual nanogaps. This is straightforward; simply extract the charge values from the original 2.5D model according to the specifications of nanogap width and length, and repeat this at different positions in the 2.5D model to mimic the stochasticity and device-to-device variation in real fabrication processes.

Second, follow the perturbation strategy developed in Paper 1 and produce the band fluctuation of $a$-Si:H in individual nanogaps. This relies on the three-dimensional fast multipole method algorithm (FMM3D, developed by the Flatiron Institute [38] and adopted in Paper 1), to accelerate the calculation.

Third, produce the DOS distribution. The DOS distribution of $a$-Si:H in the entire nanogap as well as in **R** are easily obtained based on the band fluctuation result (using the same perturbation method in Paper 1). Afterwards, the division of lanes is specified according to Eqs. (6) and (18). The DOS distribution of $a$-Si:H in each lane is then similarly calculated.

Fourth, calculate the nanogap mobility. This relies on the virtual film construction in Fig. 4(b), based on which the nanogap mobility is derived using Eq. (30).

The integrations are approximated by summations using an energy interval of 0.001 eV.

Fig. 9 and 10 require extra 2.5D charge models to simulate different extent of medium-range order. Generating these charge models entirely follows the procedures detailed in Paper 1. (i) Start with the generation of Gaussian array of random charge values (with a standard deviation $\sigma$). (ii) Smooth the array using the moving average algorithm; different window dimensions ($w$) represent different medium-range orders. (iii) Calculate the band fluctuation using FMM3D algorithm. (iv) Produce the DOS distribution of the respective 2.5D models. (v) Retrospectively adjust $\sigma$ to alter the respective DOS curves till they match the DOS curve of the $w = 5a_0$ model.

The decomposition in Fig. 1 and Fig. 9 is completed using the built-in Fourier decomposition analysis of OriginPro. Focus is given to normalized power distribution, based on which the cumulative distribution is obtained. To define what spatial frequencies are "predominant", a criterium of a 95% cumulative distribution is used.

Except the FMM3D algorithm which is conducted in Fortran due to the ease of setup, all the other computations are done in MATLAB. Fittings are completed in OriginPro.